\documentclass[aps,prl,twocolumn,reprint,amsmath,amssymb,preprintnumbers,superscriptaddress,bibnotes,nofootinbib,floatfix]{revtex4-1}

\usepackage{graphicx} 
\usepackage{dcolumn} 
\usepackage{footmisc} 
\usepackage{mathtools}
\usepackage{bm} 
\usepackage[colorlinks = true, linkcolor = purple, urlcolor  = cyan, citecolor = blue, anchorcolor = blue]{hyperref}
\usepackage{xcolor}
\usepackage{float}
\usepackage{amsmath}

\newcommand{\prlsection}[1]{{\it\textbf{#1}.}---}

\begin{document}  

\preprint{KEK-QUP-2025-0022, KEK-TH-2767, KEK-Cosmo-0394} 

\title{Probing dark matter interactions with stellar motion near Sagittarius A*}

\author{R. Andrew Gustafson}

\affiliation{
Center for Neutrino Physics, Department of Physics, Virginia Tech, Blacksburg, VA 24061, USA}
\affiliation{
International Center for Quantum-field Measurement Systems for Studies of the Universe and Particles (QUP,WPI), High Energy Accelerator Research Organization (KEK), Oho 1-1, Tsukuba, Ibaraki 305-0801, Japan}

\author{Ian M. Shoemaker}
\affiliation{
Center for Neutrino Physics, Department of Physics, Virginia Tech, Blacksburg, VA 24061, USA}

\author{Volodymyr Takhistov}

\affiliation{
International Center for Quantum-field Measurement Systems for Studies of the Universe and Particles (QUP,WPI), High Energy Accelerator Research Organization (KEK), Oho 1-1, Tsukuba, Ibaraki 305-0801, Japan}
\affiliation{Theory Center, Institute of Particle and Nuclear Studies (IPNS), High Energy Accelerator Research Organization (KEK), Oho 1-1, Tsukuba, Ibaraki 305-0801, Japan}
\affiliation{
 Graduate University for Advanced Studies (SOKENDAI), Oho 1-1, Tsukuba, Ibaraki 305-0801, Japan}
 \affiliation{ Kavli Institute for the Physics and Mathematics of the Universe (WPI), The University of Tokyo Institutes for Advanced Study, The University of Tokyo, Kashiwa, Chiba 277-8583, Japan
}

\begin{abstract}
Stars orbiting Sagittarius A* (Sgr A*) at the Milky Way's center provide a unique laboratory to test gravity and dark matter (DM). We demonstrate that DM interactions in stellar interiors induce a novel momentum transfer force, altering orbits beyond gravitational effects. We derive the first astrophysical constraints on DM-nucleon scattering using S2's 2000-2019 orbital data probing sub-GeV parameter space for specified DM distributions near Sgr A*. We further show that complementary orbital survival constraints and future observations of highly eccentric S-stars extend the reach to DM profiles allowed by present gravitational measurements.
Unlike bounds requiring DM capture and annihilation, our constraints apply broadly to scattering DM  including asymmetric DM, and extend to masses where methods based on capture can  lose sensitivity. Stellar lifetime constraints over Myr timescales complement these, surpassing some direct detection and cosmological limits.  This work establishes stellar dynamics as a novel probe of DM interactions.  
\end{abstract}
 
\maketitle 

\prlsection{Introduction} 
The Milky Way's Galactic Center provides a unique natural laboratory for testing gravity and probing the fundamental properties of dark matter (DM), the Universe's dominant matter, detected only gravitationally thus far.  
Observations of S-cluster stars orbiting the Galactic Center have confirmed the presence of a supermassive black hole Sagittarius A* (Sgr A*) with mass of $M_{\rm BH} \simeq 4 \times 10^{6} M_{\odot}$ at $\sim 8$ kpc ~\cite{Ghez:1998ph,Eckart:1996zz}. Precision measurements of stellar motion, particularly S2, have enabled stringent tests of general relativity~\cite{S2OrbitData}. 

DM is expected to cluster around supermassive black holes due to gravity, forming overdense ``spikes'' that can enhance the local DM density by orders of magnitude \cite{peebles1972gravitational,quinlan1994models,Gondolo:1999ef}. These overdensities can significantly affect DM observations and signatures~\cite{Gondolo:1999ef,Gnedin:2003rj,Bertone:2002je,Nishikawa:2017chy,Ding:2024mro,S2Ultralight}. A variety of studies have explored the impact on the S-cluster stars~\cite{GRAVITY:2024tth,Tomaselli:2025zdo,SgrAScatter}. 

Here we put forth a novel approach for probing fundamental DM interactions by considering scattering of DM particles with constituents inside stellar interiors
that induce a net momentum transfer capable of altering stellar orbital motion and, over extended timescales, degrading the orbits.
These effects depend simultaneously on the macroscopic DM distribution near Sgr A* and the microscopic DM scattering cross sections. 
Previous studies examined effects such as stellar heating or long-range forces. In contrast, we demonstrate that momentum transfer from DM  scattering with Standard Model particles produces a direct nongravitational force that modifies orbital dynamics, enabling new constraints such as from astrometric measurements and radial velocities inferred from spectroscopy, hereafter jointly referred to as ``astrometric measurements". Thus, our approach provides a new bridge between stellar dynamics and dark sector particle physics. 

Unlike stellar probes based on DM capture followed by annihilation or energy deposition, the  mechanism developed here requires neither capture nor annihilation. It thus applies to asymmetric and other non-annihilating DM scenarios and remains effective at sub-GeV masses where methods based on capture can lose sensitivity. Our observables are also distinct from purely gravitational drag or long-range force effects as it directly probes short range microscopic scattering between DM and Standard Model particles and maps the resulting orbital signatures onto the particle interaction properties used by laboratory searches.

\begin{table*}[t]
\centering
$
\begin{array}{r|cccccccc}
\hline\hline
\text{Stars} & M_{s} (M_{\odot}) & R_{s} (R_{\odot}) & a \mathrm{(AU)} & e & T_{0} (\mathrm{yr}) & i({}^{\circ}) & \omega ({}^{\circ}) & \Omega ({}^{\circ})\\
\hline
\text{S2}    & 13.6^{+2.2}_{-1.8} & 5.53^{+1.77}_{-0.79} & 1007 \pm 1.5 & 0.886 \pm 0.0004 & 2018.37 \pm 0.0004 & 133.88 \pm 0.18 & 66.03 \pm 0.24 & 227.40 \pm 0.29\\
\text{S62}   & 6.1^{+2}_{-1}      & 4.24^{+1.08}_{-1.16} & 737 \pm 4     & 0.976 \pm 0.01    & 2003.33 \pm 0.02   & 72.76 \pm 5.15   & 42.62 \pm 2.29  & 122.61 \pm 4.01\\
\text{S4711} & 2.2^{+2}_{-1}      & 1.87^{+1.27}_{-0.72} & 616 \pm 12    & 0.768 \pm 0.03    & 2010.85 \pm 0.06   & 114.71 \pm 2.92  & 131.59 \pm 3.09 & 20.10 \pm 3.72\\
\text{S4714} & 2.0^{+2}_{-1}      & 1.74^{+1.29}_{-0.74} & 837 \pm 2     & 0.985 \pm 0.011   & 2017.29 \pm 0.02   & 127.70 \pm 0.28  & 357.25 \pm 0.80 & 129.28 \pm 0.63\\
\hline\hline
\end{array}
$
    \caption{Parameters of the S-cluster stars analyzed in this work. Values for S2 are taken from Ref.~\cite{habibi2017twelve, S2OrbitData}, and those for S62, S4711, and S4714 from Ref.~\cite{peissker2020s62}. For stars other than S2, the radius is estimated as $R_{s} = R_{\odot} (M_{s}/M_{\odot})^{0.8}$. The variables $i$, $\omega$, and $\Omega$ denote the inclination, longitude of the pericenter  and the position angle of the ascending node, respectively. $T_{0}$ is the time of closest approach to Sgr A*. 
    \label{tab-star-parameters}}
\label{tab-star-parameters}
\end{table*}

\prlsection{Dark matter properties}
We denote the DM particle by $\chi$ with mass $m_{\chi}$ and characterize its density profile near the Galactic Center with a power-law spike~\cite{Gondolo:1999ef}
\begin{equation}
\rho_{\chi}(r) = \rho_{\mathrm{sp}} \left(\frac{R_{\mathrm{sp}}}{r}\right)^{\gamma}~,
\end{equation}
where $\rho_{\mathrm{sp}}$ is the density at the outer spike boundary, $\gamma$ is the index, and $R_{\rm sp}$ is the spike radius. Although the DM profile is expected to be modified within a few Schwarzschild radii of supermassive black holes due to relativistic effects~\cite{Sadeghian:2013laa,Speeney:2022ryg},  and at distant $r \gg R_{\rm sp}$ regions the DM halo is expected to follow a characteristic distribution~\cite{Navarro:1995iw},  the description is valid for the stellar orbit analysis considered here. While the exact structure and evolution of the DM spike depend on the Galactic Center's dynamical history, the adopted power-law form provides a well-motivated representative benchmark. As in other astrophysical DM constraints, the exact quantitative sensitivity depends on the underlying DM distribution. 

The DM velocity $\mathbf{v_{\chi}}$ distribution is considered to be a Maxwell-Boltzmann distribution in the supermassive black hole frame. In the star's frame with velocity $\mathbf{v_s}$ one has
\begin{equation}
    f(\mathbf{v_{\chi}}) = \frac{1}{v_{\rm char}^3 \pi^{3/2}} \exp\bigg(-\frac{(\mathbf{v_{\chi}} + \mathbf{v_{s}})^2}{v_{\rm char}^2} \bigg)
    \label{eq-vel-dist}
\end{equation}
where $v_{\rm char} = \sqrt{ 2 G M_{\rm BH} / \mathcal{F} r } $, $G$ the gravitational constant, and the factor $\mathcal{F}$ ranges from 2.8 to 5.5 in analyses of active galactic nuclei \cite{AGN_Dispersion:2004,AGN_Dispersion:2013,AGN_Dispersion:2014}. We adopt $\mathcal{F}=3.5$, finding that the results are not significantly affected by this choice.

For DM interactions with Standard Model particles we focus on minimal spin-independent scattering off nucleons, with our analysis methodology readily applicable to other DM interactions.
The temperature of the considered stars is sufficiently low such that we can treat the nucleons as at rest relative to DM. The differential cross section with respect to nucleon recoil energy $E_R$, and momentum transfer $q = \sqrt{2 m_{N} E_{R}}$ for nucleus of mass $m_N$, is 
\begin{equation}
    \frac{d \sigma}{dE_{R}} = \frac{\bar{\sigma}_{n} m_N}{2 \mu_n^2 v_{\chi}^2}  \bigg(\frac{m_{V}^2 + q_{0}^2}{m_{V}^2+q^2} \bigg)^2 \big|F_{N}(q) \big|^2
\end{equation}
where $m_{V}$ is the mediator mass, 
$\mu_{n}$ is the DM-nucleon reduced mass, $\bar{\sigma}_{n}$ is a characteristic cross section and $q_{0} = \mu_{n} v_{\chi}$. Here we work in natural units. For considered energies the nuclear form factor is always $F_{N}(q) = 1$ \cite{ANGELI2004185}.

We define
\begin{equation}
    \sigma_{T} = \int \frac{d\sigma}{dE_{R}} \frac{m_{N} E_{R}}{\mu^2 v_{\chi}^2}dE_{R} \label{eq-sigma-T} 
\end{equation}
and the optical depth analog  $\tau_{T} = 2 R_{s} n_{N} \sigma_{T} m_{N}/(m_{\chi} + m_{N}) $ for star of mass $M_s$ and radius $R_s$ and nucleon number density $n_{N} = (M_{s}/m_{N}) (4\pi R_{s}^{3}/3)^{-1}$. For a DM particle with velocity $\mathbf{v_\chi}$ traversing the star, the average momentum imparted to the star is
\begin{equation}
\langle \mathbf{p_{SM}} \rangle  = m_{\chi} \mathbf{v_{\chi}} \bigg(1-2 \frac{1 -\exp(-\tau_{T}) (1 +\tau_{T})}{\tau_{T}^2} \bigg)~, \label{eq-psm}
\end{equation}
with derivation found in Supplemental Material A.

We find the force on the star from DM interactions is
\begin{equation}
    \mathbf{F_{scat}} = \pi R_{s}^2 \frac{\rho_{\chi}}{m_{\chi}} \int v_{\chi}  \langle \mathbf{p_{SM}}\rangle f(\mathbf{v_{\chi}}) d^{3}\mathbf{v_{\chi}} \label{eq-star-force}
\end{equation}
where $v_{\chi} = |\mathbf{v_{\chi}}|$. Equation~\eqref{eq-psm} saturates to $m_\chi \mathbf{v_\chi}$ for $\tau_T \gg 1$, yielding a maximal force scaling linearly with $\rho_\chi$ and nearly quadratically with $v_s$ when $v_s \gg v_{\rm char}$. Thus, in the saturated regime the characteristic scaling is  $F_{\rm scat}\sim\pi R_s^2\rho_\chi v_s^2$ demonstrating that stars on tighter and high velocity orbits through dense DM environments are generically favored as probes. Our general mechanism is therefore not specific to S2 or to the Milky Way benchmarks and can be applied to other S-stars and, more generally, to stars orbiting supermassive black holes.

\prlsection{Effects on astrometric observations}
Precision measurements of S-cluster star motion, especially that of S2, provide sensitive probes of new physics  including tests of general relativity~\cite{S2OrbitData}. We analyze the stars S2, S62, S4711 and S4714 whose relevant parameters are described in Table~\ref{tab-star-parameters} and adopt Sgr A*
properties from Ref.~\cite{S2OrbitData}.
The principal observables are the stellar right ascension (RA), declination (Dec), and radial velocity (RV).

\begin{figure*}[t]
    \centering
    \includegraphics[width = 0.49 \textwidth]{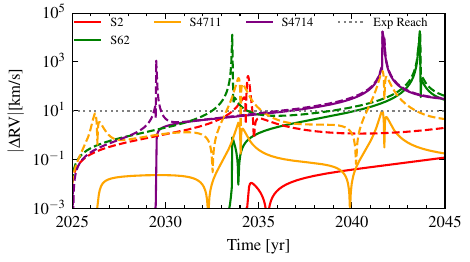}
    \includegraphics[width = 0.49 \textwidth]{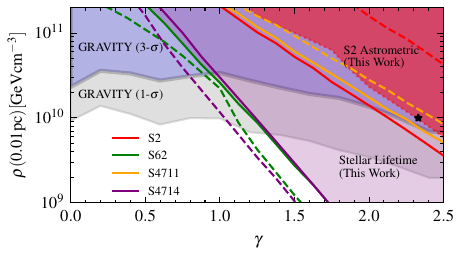}
    \caption{\textbf{Left:} RV deviations induced by DM obtained from our simulations. Dashed lines indicate the difference between a stellar orbit without a DM overdensity spike and one with a spike and only gravitational interactions. Solid lines indicate the difference between two orbits with DM spikes with and without DM-nucleon scattering. We consider a characteristic DM spike profile with $\gamma = 7/3$, $\rho(0.01 \mathrm{pc}) = 10^{10}~\mathrm{GeV ~cm^{-3}}$ as well as $\tau_{T} \gg 1$ for DM scattering. The dotted horizontal line indicates the reach of 10~km~s$^{-1}$ for near future astrometric measurements \cite{S2OrbitData,GRAVITY-Astrometric,Astrometry30MT,NIFS-Parameters,Osiris-Resolution}.
   \textbf{Right:} DM density at 0.01 pc versus spike index for $m_{\chi} = 10^{-2}$~GeV and $\sigma_n =   5 \times 10^{-35}~ \mathrm{cm^{2}}$ with a heavy mediator. Solid lines denote tidal disruption events within 1 Myr, the purple shaded region indicates parameters where any of the four stars would undergo a tidal disruption. Dashed lines correspond to observational sensitivities of 10~km~s$^{-1}$ or 0.1 mas.  The red region denotes where DM scattering effects on S2 in 2000-2019 data cause deviations exceeding measurement uncertainties~\cite{S2OrbitData}. 
   The 1-$\sigma$ and 3-$\sigma$ gravitational only exclusions from GRAVITY~\cite{GRAVITY:2024tth} are shown. The black star indicates the parameters used in Fig.~\ref{fig:sigma}. 
    \label{fig-Direct-Observables}}
\end{figure*}

DM can influence these stellar orbits through three distinct mechanisms.  Purely gravitational effects arise from extended mass distribution~\cite{GRAVITY:2024tth} and also induce dynamical friction (e.g.~\cite{Tomaselli:2025zdo}, and earlier proposed in the context of other supermassive black holes~\cite{Chan:2024yht,Ding:2025nxe}). Our interest is in direct DM scattering with nucleons within stellar interiors, which produces a novel nongravitational momentum transfer force described by the formalism developed above.
Among these, only the last effect directly probes DM microphysics.

To investigate the impact of these DM effects, we perform three distinct types of first post-Newtonian order orbital simulations (see Supplemental Material B for details). Namely, we consider: (i) a reference case with only supermassive black hole, (ii) a case of supermassive black hole plus a DM spike affecting the orbit solely through gravity and (iii) a case of supermassive black hole with DM spike and including DM-nucleon interactions. We then compare the simulated astrometric observables between cases (i) and (ii), corresponding to purely gravitational effects, and between (ii) and (iii), corresponding to DM scattering interaction effects. We find that dynamical friction is negligible compared to other effects across the relevant parameter space and omit it (see Supplemental Material C and the references therein \cite{1943ApJ....97..255C} for justification).

In Fig.~\ref{fig-Direct-Observables} we display resulting differences in radial velocities considering
a large cross section ($\tau_{T} \gg 1$), with RA and Dec found in Supplemental Material B.
At early times, gravity only perturbations dominate. As the star approaches pericenter, where both its velocity and the DM density are largest, the DM scattering-induced deviations rapidly grow. We estimate that subsequent passage through the pericenter will occur in the years 2034, 2033, 2026, and 2029 for S2, S62, S4711, and S4714, respectively. For highly eccentric stars such as S62 and S4714 the DM scattering effects can exceed purely gravitational contributions. Since the two effects are nondegenerate  astrometric observations could disentangle them. 

Our results are illustrated considering reference DM spike profile of $\gamma = 7/3$, consistent with expectations~\cite{Gondolo:1999ef} from adiabatic formation from an initial Navarro-Frenk-White DM profile \cite{Navarro:1995iw}. 
For illustration, we adopt the benchmark
$\rho_{\chi}$(0.01~pc) = $10^{10} ~\mathrm{GeV~cm^{-3}}$, which lies just outside the current $3-\sigma$ GRAVITY measurement uncertainty~\cite{GRAVITY:2024tth} for the smooth power-law density profile considered here. The dependence of our results on the density normalization and slope, together with the GRAVITY constraints, is shown in Supplemental Material. 

Other values can be readily accommodated by our framework, and we examine the effect of varying spike parameters in the Supplemental Material D. We use this DM profile as an optimistic but well-motivated benchmark and this is not a prerequisite for our mechanism. The GRAVITY bound applies to the total extended mass within the object's orbit including stars, compact remnants  and DM. 
For  DM  with  an overdensity spike the lower central DM density can substantially weaken these constraints, whereas steep adiabatic spikes can maximize accessible scattering parameter space. This astrophysical dependence is complementary to the particle physics dependence encoded in the scattering cross section. \\ 
\newline
We emphasize that probes such as GRAVITY search regimes constrain the gravitational effect of the total extended mass, whereas our mechanism probes the microscopic DM-nucleon scattering cross section. The two observables are therefore physically distinct. For a fixed DM profile  gravitational measurements restrict the allowed density normalization, while the orbital scattering signal independently constrains the interaction strength.

\begin{figure*}[t]
    \centering
    \includegraphics[width = 0.49 \textwidth]{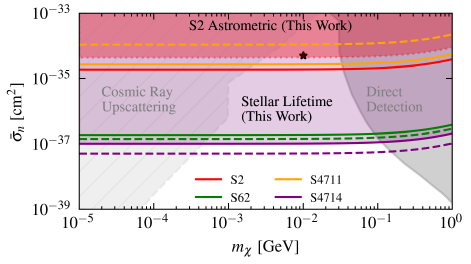}
    \includegraphics[width = 0.49 \textwidth]{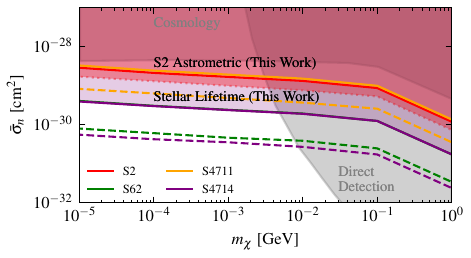}
    \caption{DM-nucleon scattering cross section versus DM mass for a heavy mediator (\textbf{left}) and a light mediator with $m_V = 1$~eV (\textbf{right}). A characteristic DM overdensity spike index of  $\gamma = 7/3$  and density $\rho(0.01 \mathrm{pc}) = 10^{10}~\mathrm{GeV ~cm^{-3}}$ is considered. The red region denotes where DM scattering effects on S2 in 2000-2019 data cause
deviations exceeding measurement uncertainties~\cite{S2OrbitData}.
In deriving these constraints the stellar orbital elements are held at their measured best fit values and the DM profile parameters are set. The scattering  deviation is sharply peaked at pericenter while gravitational perturbations from the extended mass accumulate smoothly; hence the two are not degenerate in their time dependence. Solid lines mark tidal disruption events within 1 Myr, the purple region indicating parameters where any of the four stars would undergo a tidal disruption. Dashed lines correspond to deviations of 10 km~s$^{-1}$ in RV or 100 $\mu$as in RA or Dec over 20 yr period, comparable to current and near future sensitivities~\cite{S2OrbitData, GRAVITY-Astrometric,Astrometry30MT,NIFS-Parameters,Osiris-Resolution}. Limits from direct detection from SENSEI, XENON, LUX, and PandaX \cite{SENSEI:2020dpa,SENSEI:2023zdf,Migdal2019,LUX:2018akb,PandaX-II:2018xpz} as well as cosmological constraints \cite{Buen-Abad:2021mvc} that include CMB with Baryon Acoustic Oscillations, Lyman-$\alpha$ forest  and Milky Way satellite measurements are shown. We also display cosmic-ray boosted DM bound estimates for IceCube \cite{Cappiello:2024acu} and limits from Super-Kamiokande \cite{Super-Kamiokande:2022ncz}, indicated with hatched contours and dashed lines reflecting their additional model dependence and complementarity. The black star in the left panel marks parameters used in Fig.~\ref{fig-Direct-Observables}.
    \label{fig:sigma}}
\end{figure*}

Using 19 years of existing S2 observational data~\cite{S2OrbitData}
we estimate novel exclusions for DM interactions. Fixing the DM spike index  $\gamma = 7/3$  and density at 0.01 pc we vary the DM mass and scattering cross section. We simulate the S2 motion from 2000 to 2019 with and without scattering (see Supplemental Material B). The average observational 1-$\sigma$ uncertainties during this period were $\simeq 550~\mu$as in RA/Dec and $\simeq 30$~km~s$^{-1}$ in RV. If the scattering effect alone produces deviations of this magnitude  we consider the corresponding parameter point excluded and in the red region denoted ``S2 Astrometric'' of  Fig.~\ref{fig:sigma}. 

For the smooth power-law DM density profiles considered here imposing the current GRAVITY $3-\sigma$ constraint substantially reduces the parameter space accessible with existing S2 data alone, with the reach approaching the GRAVITY boundary only for the steepest density profiles. This illustrates the strong dependence of the present numerical constraints on the DM profile. We emphasize that our underlying scattering mechanism is also applicable to other stellar systems. More eccentric stars with smaller pericenters, such as S62 and S4714, provide substantially enhanced reach because the scattering effect peaks where both the stellar velocity and DM density are largest.
Our study thus calls for dedicated analyses, including using a simultaneous multiparameter fit varying the orbital, reference frame  and DM profile parameters, by experimental collaborations that can further improve on our results.

Near future experiments will be able to probe DM interactions with significantly enhanced sensitivity.
We project their reach
by considering that future astrometric precision of $\simeq 100~\mu$as in RA/Dec and $\simeq10$~km~s$^{-1}$ in RV, consistent with GRAVITY~\cite{GRAVITY-Astrometric}, the Thirty Meter Telescope~\cite{Astrometry30MT}, NIFS~\cite{NIFS-Parameters}  and OSIRIS~\cite{Osiris-Resolution}. We consider DM parameter points yielding such deviations within a 20-year observational period as detectable.  

As shown in Fig.~\ref{fig:sigma}, the exclusion limits we estimated for S2 considering 2000-2019 data exceed  the expected sensitivity (dashed lines) from future observations of S2 between 2025 and 2045. While the upcoming observations carry improved precision, S2 passed through pericenter twice during the 2000-2019 period but will do so only once in the 2025-2045 period. Since DM scattering effects peak at the pericenter, stars with larger eccentricities and smaller pericenter distances provide the strongest constraints on DM interactions. This highlights the significance of multidecade observations of these stars, as each pass through the pericenter leads to compounding effects.

\prlsection{Effects on long term  orbital survival}
DM can impact long-term survival of stars by inducing orbital energy losses and decay. While minuscule on human timescales, dynamical processes near supermassive black holes, such as relaxation, are expected to act over $\mathcal{O}(10^{6})$ yr \cite{StellarDynSMBH}.  
Thus, we impose new indirect limits on DM interactions by requiring that stellar orbits remain stable for at least this timescale. These constraints are conditional on the star having occupied  or repeatedly sampled  an orbit comparable to its present one over this interval. Recent injection or replenishment onto the observed orbit could weaken this interpretation and would require a population-level analysis of the nuclear star cluster. This qualification does not affect the independent constraint derived from the 19-year S2 data set. 

We note that stellar relaxation~\cite{StellarDynSMBH} is not expected to erase the orbital energy loss considered here on the relevant timescales. At the semimajor axes of the observed S-stars  the fastest vector resonant relaxation conserves orbital energy and the magnitude of angular momentum, while nonresonant relaxation  process that diffuses orbital energy occurs on much longer timescales. The million-year survival requirement therefore isolates DM-induced dissipation under the stated orbital history assumption. For sufficiently steep DM density profiles  this orbital survival constraint extends into the region allowed by the current GRAVITY observations at $3-
\sigma$ uncertainty level, as shown in Supplemental Material Fig. 6.

Among the possible mechanisms both dynamical friction and DM scattering lead to energy losses. A static extended mass distribution only modifies the gravitational potential without dissipating energy.
Thus, DM scattering processes are dominant contributions to stellar orbital evolution in our analysis.

We define the end of the stellar lifetime when the star crosses its tidal disruption event radius  $r_{\rm TDE} = (M_{\rm BH}/M_{s})^{1/3} R_{s}$. Due to slowly varying changes we average the evolution over many Keplerian orbits for which the orbital position and velocity depend on the true anomaly $f$, semimajor axis $a$, and eccentricity $e$ as
\begin{equation}
    r(f) =  \frac{a(1-e^{2})}{(1+e \cos(f))} ~~~,~~~ \frac{df}{dt} = \frac{n\big( 1+e\cos(f) \big)^2}{(1-e^2)^{3/2}}~ ,
\end{equation}
with $n = \sqrt{G M_{\rm BH}/a^{3}}$. Following Refs.~\cite{Ecc_Semi_Major_Evolution,Dosopoulou:2023umg}, we introduce a dimensionless function $\epsilon(r,v_{s})$ such that
\begin{equation}
    \epsilon(r,v_{s}) \mathbf{v_{s}} = - \frac{\mathbf{F} v_{s}^3}{M_{s}}.
    \label{eq-epsilon}
\end{equation}
where $\mathbf{F}$ is the total force acting on the star.

Averaging over many orbital periods yields parameter evolution equations~\cite{Dosopoulou:2023umg,Ecc_Semi_Major_Evolution}

\begin{widetext}

\begin{equation}
    \bigg \langle \frac{da}{dt} \bigg \rangle = \frac{(1-e^2)^2}{\pi n^3 a^2} \int_{0}^{2\pi} \frac{(1+e \cos(f))^{-2} \epsilon(r,v_{s})}{(1+e^2 + 2 e \cos(f))^{1/2}} df~,
    \label{eq-semimajor-evolution}
\end{equation}

\begin{equation}
    \bigg \langle \frac{de}{dt} \bigg \rangle = \frac{(1-e^2)^3}{\pi n^3 a^3} \int_{0}^{2\pi} \frac{(e+\cos(f)) \epsilon(r,v_{s})}{(1+e^2 + 2 e \cos(f))^{3/2} (1+e \cos(f))^2} df~.
    \label{eq-eccentricity-evolution}
\end{equation}

\end{widetext}

Using the present day orbital parameters in Table~\ref{tab-star-parameters} we determine the orbital evolution. We exclude
DM parameters that result in tidal disruption events within 1 Myr, shorter than the stellar lifetime~\cite{habibi2017twelve} or other dynamical processes near Sgr A*~\cite{StellarDynSMBH}. This constrains the DM profile overdensity and slope for a given DM mass and cross section, as shown in Fig.~\ref{fig-Direct-Observables}. Equivalently, this restricts the DM interaction parameters for fixed DM overdensity profile properties as shown on Fig.~\ref{fig:sigma} by the purple ``Stellar Lifetime'' region that denotes the parameter space where any of the four stars S2, S62, S4711, S4714 would be disrupted within 1 Myr.

The constraint's dependence  on the scattering optical depth $\tau_{T}$, as used in Eq. \eqref{eq-psm}, illustrates effects of imposing alternative stellar lifetime restrictions.  If $\tau_{T} \ll 1$, then stellar lifetime scales as $\tau_{T}^{-1} \propto \sigma_{n}^{-1}$, while for  $\tau_{T} \gtrsim 1$  the stellar lifetime saturates and further increases in the cross section do not shorten the time to disruption.

\prlsection{Conclusions}
We have established a previously overlooked mechanism through which DM interactions within stellar interiors induce a net momentum transfer that can significantly alter stellar motion around the Galactic Center supermassive black hole Sgr A*. This effect directly links the microphysics of DM interactions to the macroscopic orbital dynamics of stars and can exceed purely gravitational perturbations. 
Applying this framework to the decades-long orbital observations of S2  we derive the first astrophysical constraints on DM-nucleon scattering from stellar orbital data for specified Galactic Center DM density profiles. The quantitative reach depends on the DM distribution, while complementary orbital survival constraints and future observations of highly eccentric S-stars extend sensitivity into profile parameter space allowed by present gravitational measurements. 
The predicted signatures depend on orbital eccentricity and pericenter distances, making eccentric S-cluster stars the most sensitive probes of DM overdensities near the black hole.
Requiring stellar survival over Myr-timescales yields  independent and complementary constraints exceeding cosmological and direct detection bounds by orders of magnitude for considered DM distributions. We found that high velocity stars with small pericenters in dense DM environments are particularly sensitive probes. The framework therefore applies to a broad variety of DM distributions, stars  and stellar systems around supermassive black holes.   Using our approach, multiparameter fits and  stellar population considerations will provide additional handle on DM searches.
 
Our results establish stellar orbital dynamics as a new laboratory for   probing dark sector microphysics and interactions, and motivates future high precision astrometric campaigns that will probe DM microphysics in extreme gravitational environments.

~\newline 
\textit{Note added:---} During the final stages of this work, a preprint appeared~\cite{Acevedo:2025rqu} that also considered DM  scattering effects and impact on Galactic Center orbits.
Our analysis offers several essential advances and provides a complementary perspective:
we derive the first constraints based on existing decades-long S2 astrometric data, perform full orbital simulations including scattering dynamics, present 20 year observational projections  and develop a novel formalism for long-term evolution. 

\textit{Acknowledgments.---} We are grateful to Wick Haxton and Gonzalo Herrera for useful discussions.
R.A.G. and I.M.S. are supported by the U.S. Department of Energy under the award number DE-SC0020262. R.A.G. is partially supported by the World Premier International Research Center Initiative (WPI), MEXT, Japan. R.A.G. is grateful to QUP for hospitality during the visit in which this project was initiated.
V.T. acknowledges support by the World Premier International Research Center Initiative (WPI), MEXT, Japan and JSPS KAKENHI grant No. 23K13109.
This material is based upon work supported by the U.S. Department of Energy, Office of Science, Office of Workforce Development for
Teachers and Scientists, Office of Science Graduate Student Research (SCGSR) program. The SCGSR program is administered by the
Oak Ridge Institute for Science and Education (ORISE) for the DOE. ORISE is managed by ORAU under contract number
DESC0014664. R.A.G. is a recipient of the SCGSR Award.

\bibliography{main}
\phantom{i} 

\appendix
\newpage

\onecolumngrid
\pagebreak
\centerline{\large {Supplemental Material for}}
\medskip

{\centerline{\large \bf{Probing Dark Matter Interactions with Stellar Motion near Sagittarius A*}}}
\medskip
{\centerline{R. Andrew Gustafson, Ian M. Shoemaker, Volodymyr Takhistov}}
\bigskip
\bigskip

In this Supplemental Material we provide additional details on the derivation of the momentum transfer formalism for DM-nucleon scattering, the description of orbital simulations and observables, the comparative analysis of other orbital energy loss processes  and supplementary figures illustrating DM-spike parameters and stellar-orbit evolution.

\section{A. Dark Matter Scattering Momentum Transfer Formalism}

We detail here the derivation of the momentum transfer induced by DM-nucleon scattering. Consider a nucleon initially at rest and a DM particle with non-relativistic velocity $v_{\chi}$. After scattering  the nucleon recoils with energy $E_{R}$ and the outgoing DM is deflected by an angle $\theta$. From kinematics  one has
\begin{equation}
     m_{N} E_{R} = \big( E_{i} E_{f} -p_{i} p_{f} \cos \theta - m_{\chi}^2 \big)
\end{equation}
where $E_{i}$ and $E_{f} = E_{i} - E_{R}$ is the initial and final DM particle energies with corresponding initial momentum $p_{i} = \sqrt{E_{i}^2 - m_{\chi}^2}$ and analogously a final momentum $p_f$. In the non-relativistic limit
\begin{equation}
    \frac{p_{f} \cos \theta}{p_{i}} \simeq 1 - \frac{E_{R} (m_{N} + m_{\chi})}{m_{\chi}^2 v_{\chi}^2}.
\end{equation}
Thus, the average fraction of momentum retained in the initial direction after a single scattering is
\begin{equation}
    \bigg \langle \dfrac{p_{f} \cos \theta}{p_{i}}  \bigg \rangle = 1 - \frac{m_{N}}{(m_{\chi} + m_{N})} \dfrac{\sigma_{T}}{\sigma} \label{eq-frac-mom},
\end{equation}
where $\sigma$ is the total cross section and $\sigma_{T}$ is defined in Eq.~\eqref{eq-sigma-T}. 

We consider that scattering does not significantly alter the DM speed, such that $\sigma_{T}$ remains constant along the particle's path. This  approximation remains accurate until $m_{\chi} \simeq m_{N}$, where direct detection limits dominate. The average momentum imparted to the star is then
\begin{equation}
    \langle \mathbf{p_{SM}}\rangle = m_{\chi} \mathbf{v_{\chi}} \bigg[1 - \sum_{k=0}^{\infty} p(k)\bigg(1 - \frac{m_{N}}{(m_{\chi} + m_{N})} \frac{\sigma_{T}}{\sigma} \bigg)^{k} \bigg] \label{eq-psm-calc},
\end{equation}
where $p(k)$ is the probability that the DM undergoes $k$ scatterings while traversing the star.

Assuming a uniform nucleon number density $n_N$ and straight line DM trajectories, a conservative choice that underestimates the rates, we obtain
\begin{equation}
    p(k) = \int_{0}^{R_{s}}dR \frac{2R}{ R_{s}^2} \frac{\exp\big(-2n_{N} \sigma \sqrt{R_{s}^2 - R^{2}}  \big)(2n_{N} \sigma \sqrt{R_{s}^2 - R^{2}})^{k} }{k!}~.
\end{equation}
Substituting into Eq.~\eqref{eq-psm-calc} and summing over $k$
yields
\begin{equation}
    \langle \mathbf{p_{SM}} \rangle = m_{\chi} \mathbf{v_{\chi}} \bigg[ 1 - \int_{0}^{R_{s}} dR \frac{2R}{ R_{s}^{2}} \exp\bigg(-\frac{2m_{N}}{m_{\chi} + m_{N}} n_{N}\sigma_{T} \sqrt{R_{S}^{2} - R^{2} }\bigg) \bigg].
\end{equation}

Employing optical depth analog $\tau_{T} = 2 R_{s} n_{N} \sigma_{T} m_{N}/(m_{\chi} + m_{N})$ as in the main text, we obtain
\begin{equation}
    \langle \mathbf{p_{SM}} \rangle = m_{\chi} \mathbf{v_{\chi}} \bigg[ 1 - \int_{0}^{R_{s}} dR \frac{2R}{ R_{s}^{2}} \exp\bigg(-\tau_{T} \sqrt{1 - \dfrac{R^{2}}{R_{s}^2} }\bigg) \bigg].
\end{equation}
Evaluating the integral gives the expression of the main text
\begin{equation}
\langle \mathbf{p_{SM}} \rangle  = m_{\chi} \mathbf{v_{\chi}} \bigg(1-2 \frac{1 -\exp(-\tau_{T}) (1 +\tau_{T})}{\tau_{T}^2} \bigg)~.
\end{equation}

\begin{figure*}[t]
    \includegraphics[width = 0.49 \textwidth]{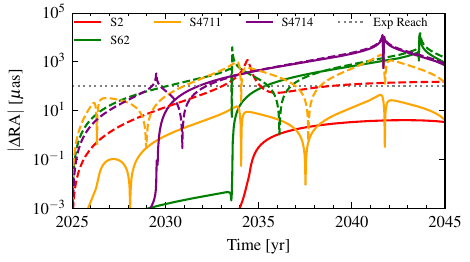}
    \includegraphics[width = 0.49 \textwidth]{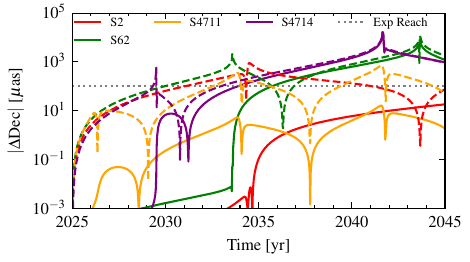}
    \caption{Same as the left panel of Fig.~\ref{fig-Direct-Observables}, but showing the differences in RA (\textbf{left}) and Dec (\textbf{right}). A characteristic DM overdensity spike index of $\gamma = 7/3$ and density $\rho(0.01 \mathrm{pc}) = 10^{10}~\mathrm{GeV ~cm^{-3}}$ is considered. Dashed lines denote purely gravitational effects, while solid lines include DM induced scattering. \label{fig-RA-Dec-Diff}}
\end{figure*}

\section{B. Simulations of Stellar Orbits with Dark Matter Interactions}

Our orbital simulations build upon the framework  outlined in Ref.~\cite{S2OrbitData}, with key modifications to incorporate DM scattering effects.

The stellar motion is computed using the first order post-Newtonian expansion, with $M_{\rm enc}(r)$ denoting the total enclosed mass within radius $r$. We account for both the supermassive black hole and the DM spike. We introduce an additional nongravitational force term $\mathbf{F_{scat}}$
representing the DM-nucleon scattering momentum transfer effects derived in Sec. A. The full equation of motion is
\begin{equation}
    \frac{d^2\mathbf{r}}{dt^2} = - \frac{G M_{\rm enc}(r) \mathbf{r}}{r^{3}} + \frac{G M_{\rm enc}(r)}{c^{2} r^{3}} \bigg( \frac{4 G M_{\rm enc}(r)}{r} - v^{2} \bigg) \mathbf{r} + \frac{4 G M_{\rm enc}(r) (\mathbf{r} \cdot \mathbf{v})}{c^{2}r^{3}} \mathbf{v} + \frac{\mathbf{F_{scat}}}{M_{s}},
    \end{equation} 
where we have explicitly added back in factors of speed of light $c$ for clarity in this section. We obtain the initial conditions using the stellar parameters in Table~\ref{tab-star-parameters}.
When modeling observables, we include the R\"omer time delay from light propagation 
\begin{equation}
    t_{\rm em} \simeq t_{\rm obs} - \frac{Z(t_{\rm obs})}{c},
\end{equation}
where $t_{\rm em}$ is the emission time corresponding to the observed time  $t_{\rm obs}$ and line-of-sight distance $Z(t)$. An iterative approach can further refine this relation.
The emission time is then used to determine the astrometric variables.

The observed RV derived spectroscopically includes both special-relativistic and general-relativistic corrections. It relates to the true stellar velocity components ($V_X, V_Y, V_Z)$, defined in the observer's coordinate frame  with $Z$ along the line-of-sight, as
\begin{equation} 
    {\rm RV}(t_{\rm obs})  = V_{Z}(t_{\rm em}) + \frac{V_{X}^{2}(t_{\rm em}) + V_{Y}^{2}(t_{\rm em}) + V_{Z}^{2}(t_{\rm em})}{2 c}  + \frac{G M_{\rm enc}(r)}{c r(t_{\rm em})}~.
\end{equation}
Small potential offsets in Sgr A*'s position or velocity have negligible effect on differential comparisons between simulated orbits we consider. 

\begin{figure*}[t]
    \includegraphics[width = 0.49 \textwidth]{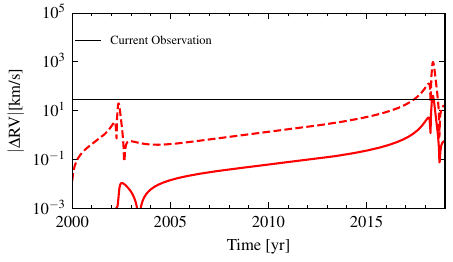}
    \includegraphics[width = 0.49 \textwidth]{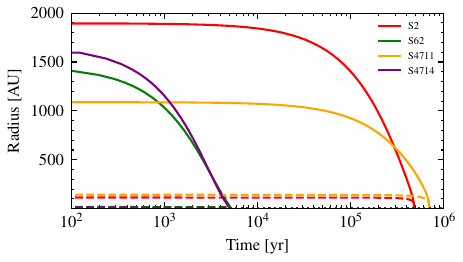}
    \caption{\textbf{Left:} Graph of the gravitational (dashed line) and scattering induced effects (solid line) on S2 radial velocity between 2000 and 2019. The horizontal line is the average uncertainty on the radial velocity measurements during this time from \cite{S2OrbitData}. \textbf{Right:} Range of the stellar radii as a function of time considering the energy loss due to scattering of DM within the spike. A characteristic DM overdensity spike index of $\gamma = 7/3$ and density $\rho(0.01 \mathrm{pc}) = 10^{10}~\mathrm{GeV ~cm^{-3}}$ is considered. Solid lines indicate the apocenter distance, while dashed lines indicate the pericenter distance.\label{motion}
    }
\end{figure*}

We computed trajectories with and without the DM scattering term to isolate its influence.
RV differences are shown in the main text, while changes in RA and Dec appear in Fig.~\ref{fig-RA-Dec-Diff}. The DM-induced modulation of S2's RV between 2000 and 2019, used to set exclusion limits in the main text using data from Ref.~\cite{S2OrbitData}, is shown in Fig.~\ref{motion}.

\section{C. Other  Energy Loss Mechanisms}

The gravitational influence of an $r$-dependent enclosed mass can appreciably modify short term stellar observables, as depicted on Fig.~\ref{fig-Direct-Observables}. However, this process does not dissipate orbital energy and thus does not affect stellar lifetimes.
The lowest order dissipative mechanisms arising solely from gravity are dynamical friction and gravitational wave emission. As we discuss below, both are subdominant to the DM scattering-induced effects for the stellar and DM parameters we consider.
We therefore neglect them in the main analysis.

\subsection{\textit{Dynamical Friction}}

Dynamical friction describes the deceleration of a massive body moving through a background medium due to the gravitational wake it induces \cite{1943ApJ....97..255C,Dosopoulou:2023umg}. For an object of mass $m$, the dynamical friction force is given as 
\begin{equation}
    \mathbf{F}_{\rm df} = -4 \pi G^2 m \rho(r) \frac{\mathbf{v_{s}}}{v_{s}^3}  \int d^{3} \mathbf{v_{\chi}} f(\mathbf{v_\chi}) \bigg(\ln\Lambda  \Theta \big(v_{s} - v_{\chi} \big)  +  \bigg[ \ln \bigg(\frac{v_{\chi} + v_{s}}{v_{\chi} - v_{s}}\bigg) -\frac{2 v_{s}}{v_{\chi}}\bigg] \Theta(v_{\chi} - v_{s}) \Theta(v_{esc} - v_{\chi}) \bigg)
\end{equation}
where non-bold quantities denote magnitudes and $\mathbf{v_{s}}$ is the DM velocity distribution in the supermassive black hole frame where it is isotropic. The escape velocity is $v_{esc} = \sqrt{2 G M_{\rm BH}/r}$, and $\ln \Lambda \sim \mathcal{O}(10)$ is the Coulomb logarithm. We have verified that the term in square brackets is always smaller than $\ln \Lambda$ and can be ignored.

Assuming a Maxwell–Boltzmann distribution as in Eq.~\eqref{eq-vel-dist}, we can solve the force exactly as
\begin{equation}
    \mathbf{F}_{\rm df} = -4 \pi G^2 m \rho(r) \ln \Lambda \dfrac{\mathbf{v_{s}}}{v_{s}^3}  \bigg(\mathrm{erf}\bigg(\dfrac{v_{s}}{v_{\rm char}} \bigg) -\dfrac{2 v_{s} \exp \bigg(- \dfrac{v_{s}^2}{v_{\rm char}^2} \bigg)}{v_{\rm char}\sqrt{\pi}}   \bigg)~.
\end{equation}

For $v_{s} \gg v_{\rm char}$ the dynamical friction force scales as $\mathbf{F}_{\rm df} \propto v_{s}^{-2}$ in contrast to DM scattering force $\mathbf{F}_{\rm scat} \propto v_{s}^2$. This opposite scaling explains why scattering dominates for the high velocity S-cluster stars. The corresponding changes in the orbital parameters are obtained by substituting $\epsilon(r,v_s)$ from Eq.~\eqref{eq-epsilon} into Eqs.~\eqref{eq-semimajor-evolution} and~\eqref{eq-eccentricity-evolution}.

\subsection{ \textit{Gravitational Waves}}

At small pericenter distances near the supermassive black hole stars lose orbital energy through the emission of gravitational waves  leading to   changes in their orbital parameters~\cite{Dosopoulou:2023umg}.
The evolution of the semimajor axis and eccentricity is given by
\begin{equation}
    \begin{split}
        \bigg \langle \frac{da}{dt} \bigg \rangle_{\rm GW} &= \frac{-64 G^{3} m M_{\rm BH}^2}{5 c^5 a^{3} (1-e^{2})^{7/2}} \bigg( 1 + \frac{73 }{24}e^{2} +\frac{37} {96}e^{4} \bigg)~,\\
        \bigg \langle \frac{de}{dt} \bigg \rangle_{\rm GW} &= \frac{-304 G^{3} m M_{\rm BH}^{2}}{15 c^{5}a^{4}(1-e^{2})^{5/2}} \bigg(1 + \frac{121}{204}e^{4} \bigg)~.
    \end{split}
\end{equation}

\begin{figure}
    \centering
    \includegraphics[width = 0.49 \textwidth]{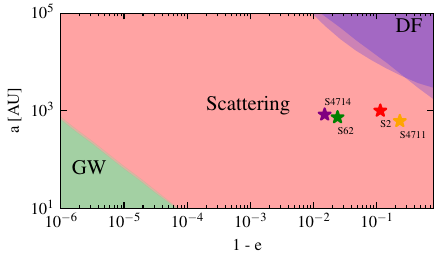}
    \caption{Dominant contribution to the orbital evolution of the semimajor axis $a$ and eccentricity $e$, as a function of $a$ and $e$, comparing DM scattering, gravitational wave (GW) emission and dynamical friction (DF). We consider DM with a mass $m_{\chi} = 10^{-2}$~GeV with a cross section of $\sigma_n = 1.37 \times 10^{-36}~\mathrm{cm^2}$. A characteristic DM overdensity spike index of $\gamma = 7/3$ and density $\rho(0.01 \mathrm{pc}) = 10^{10}~\mathrm{GeV ~cm^{-3}}$ is considered.  For this comparison we consider a star with solar mass and radius. The locations of the four S-cluster stars analyzed in this work are also indicated. \label{fig-Diff-Forces}}
\end{figure}

In Fig.~\ref{fig-Diff-Forces} we compare the relative strength of the energy loss mechanisms by considering 
the values of $a$ and $e$ for which each effect is dominant.
DM scattering effects dominate in the parameter region relevant for all four stars considered in our analysis, while gravitational wave emission becomes significant only for very small pericenters, and dynamical friction is important in the opposite, low velocity limit.
The transitions between these regimes are gradual.

In Fig.~\ref{motion}, right, we illustrate the cumulative effect of this energy loss over long timescales, showing the time evolution of the pericenter and apocenter radii induced by DM scattering.

\section{D. Dependence on Dark Matter Distribution}

\begin{figure*}
\includegraphics[width = 0.49 \textwidth]{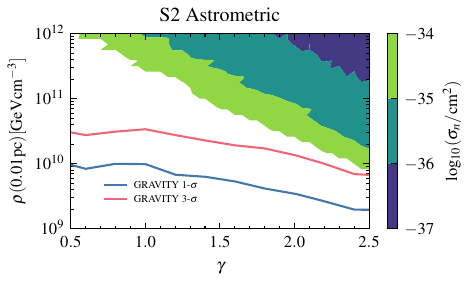}
\includegraphics[width = 0.49 \textwidth]{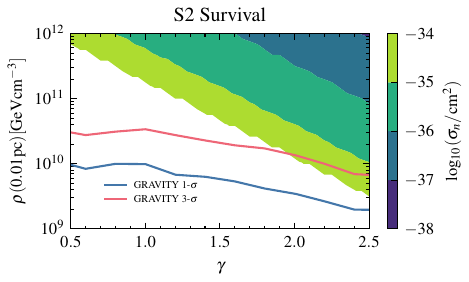}
\caption{Minimal DM nucleon-scattering cross section accessible from S2 observations between 2000 and 2019 (\textbf{left}) and from the S2 stellar lifetime constraint (\textbf{right}) as a function of the DM spike density normalization and slope.
White regions correspond to parameters for which these methods are ineffective. Also indicated are the 1- and 3-$\sigma$ constraints on the spike profile from GRAVITY observations \cite{GRAVITY:2024tth}} \label{fig-Sigma-Contour}
\end{figure*}

As discussed in the main text, the sensitivity of our analysis depends jointly on the DM microphysical interactions through the scattering cross section and the macrophysical properties of the central DM overdensity spike distribution.
Since these quantities carry some degeneracy, variations in the DM spike density profile  normalization or slope translate into different cross section values that can be probed.
Fig.~\ref{fig-Sigma-Contour} illustrates this dependence for the two types of constraints derived from S2, the 2000–2019 astrometric limit and the stellar lifetime bound. The existing S2 orbital data constraints approach  the GRAVITY $3-\sigma$ boundary only for the steepest DM density profiles considered here, whereas the complementary S2 orbital survival constraint extends into parameter space regimes allowed by GRAVITY observations for steeper profiles. This  highlights that the quantitative reach depends strongly on the  DM distribution, while the underlying scattering mechanism is independent of the specific profile choice.

Also included is a cross section-vs-DM mass plot similar to the left plot of Fig. \ref{fig:sigma}, but for a shallower spike of $\gamma = 1.5$ instead of $\gamma = 7/3$ as used in the main text. We can see that only S62 and S4714 have astrometric or lifetime sensitivities for this shallow spike, and cross sections sensitivities are much weaker than a steep spike. This further demonstrates the interplay between DM macrophysical and microphysical properties in this phenomenon.

\begin{figure}
    \centering
    \includegraphics[width = 0.49 \textwidth]{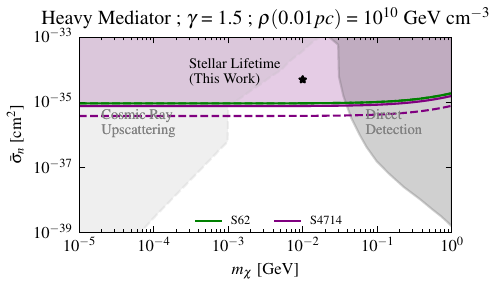}
    \caption{Cross section-vs-mass for dark matter which scatters through a heavy mediator with spike parameters $\gamma = 1.5$ and $\rho(0.01 \rm pc) = 10^{10} GeV \, cm^{-3}$. Similar to Fig. \ref{fig:sigma}, solid lines reflect stellar lifetime sensitivities while dashed lines indicate astrometric sensitivities. Only S62 and S4714 are shown, as these are the only stars for which this process can be sensitive with these spike parameters.}
\end{figure}

\end{document}